\documentstyle[aps,multicol,epsf]{revtex}

\def\beginwide{
        \end{multicols} \vspace*{-0.5cm} \noindent
        \rule{3.5in}{.1mm}\rule{.1mm}{5mm} \widetext \medskip }
\def\endwide{
        \hspace*{3.5in}~\rule[-5mm]{.1mm}{5mm}\rule{3.5in}{.1mm}
        \begin{multicols}{2}\narrowtext \vspace*{-1.0cm} \noindent }
\def\beginwidetop{
        \end{multicols} \vspace*{-0.5cm} \noindent
        \widetext \medskip }
\def\endwidebottom{
        \begin{multicols}{2} \vspace*{-1.0cm} \noindent }

\begin{document}
\bibliographystyle{unsrt}
\draft
\preprint{draft}

\title{Charge Modulation at the Surface of High--$T_{\rm c}$
  Superconductors}

\author{Thorsten Emig, Kirill Samokhin$^*$, and Stefan
  Scheidl}

\address{Institut f\"ur Theoretische Physik, Universit\"at zu
  K\"oln, Z\"ulpicher Str. 77, D-50937 K\"oln, Germany}

\date{October 1, 1997}

\maketitle

\begin{abstract}
  It is shown here that surfaces of high-temperature superconductors
  are covered by dipole layers. The charge density modulation is
  induced by the local suppression of the gap function at the surface.
  This effect is studied in the framework of the Ginzburg-Landau
  theory and crucially depends on the appropriate boundary conditions.
  Those are derived from Gor'kov's equations for a $d$-wave pairing
  symmetry. Within this framework the structure of the surface dipole
  layer is determined. The contribution of this charging to a
  lens-effect of superconducting films with holes, which has been
  studied in recent experiments, is discussed.
\end{abstract}


\begin{multicols}{2}\narrowtext 

\section{Introduction}

If superconductors were absolutely ``perfect'' conductors, they should
screen charges and would be completely free of internal electric
fields. While this is true on macroscopic length scales, it certainly
can not be expected on atomic length scales. Some recent attention
focused on the surprising fact that such electrostatic effects may
even occur on scales of the correlation length, which typically is
considerably larger than an atomic length.

Electrons in the superconductor are equilibrated if their {\em
  electrochemical} potential is spatially constant. However, the
formation of a superconducting state is in general accompanied by a
change of the chemical potential of electrons. In a spatially
inhomogeneous situation the modulation of the chemical potential
induces an accumulation of electric charge density \cite{KK,KF95,B+96}
such that the resulting electrochemical potential is constant. These
charging effects strongly depend on the ratio of the gap $\Delta$ to
the Fermi energy $\epsilon_{\rm F}$ and therefore are more strongly
pronounced in high-temperature superconductors (HTSC) than in
conventional superconductors.

Several aspects of charging effects have been examined since the
availability of the HTSC: an anomalous temperature dependence of the
work function \cite{workfun}, charge redistribution effects within the
layered structure of HTSC, and charging of vortices \cite{KF95,B+96}.
As a direct probe for the latter effect Blatter {\em et al.} suggested
\cite{B+96} to observe the electric stray field near the surface by
atomic force microscopy. Unfortunately, at present the expected
effects are beyond the resolution of this experimental technique.

In order to investigate alternative possibilities to observe such
charging effects, we examine {\em surfaces} of HTSC in a vacuum. Our
motivation is twofold. (i) The higher dimensionality of a surface
compared to a vortex line can be expected to lead to the accumulation
of much a larger charge quantities. Even if this does not necessarily
lead to much higher electric field amplitudes, the field will be
extended over a much larger region. (ii) In a recent experiment
\cite{Kri95} the influence of a thin superconducting film on an
electron beam penetrating a hole in the material has been examined. At
$T_{\rm c}$ a change of beam intensity behind the hole was observed,
i.e. the hole effectively acted as a lens. In principle, charging
induced by the suppression of the order parameter at the surface near
the hole could provide an explanation of this electro--optical effect.
This mechanism will be studied quantitatively in this work.

During the last few years increasing evidence has been found for
$d$-wave pairing instead of a conventional $s$-wave pairing in HTSC
\cite{d-wave}. For this reason we take specifically account of a
$d$-wave symmetry. In the case of a vortex line the gap vanishes for
topological reasons in the vortex center for $d$-wave pairing as well
as for $s$-wave pairing \cite{d-vortex}. The structure of the vortex
core does not feel the underlying symmetry of the order parameter, at
least in the vicinity of $T_{\rm c}$.  However, at a
superconductor/insulator interface the strength of the suppression of
$\Delta$ and eventually also the charging {\em does} crucially depend
on the symmetry.

In Section \ref{sec.bound} we give a derivation of the gap profile
near the interface for general singlet pairing on the basis of the BCS
theory. In particular, we formulate appropriate boundary conditions
for the order parameter in a phenomenological Ginzburg-Landau
description. The resulting charging effects at an infinite plane
surface are calculated in Section \ref{sec.surf}. The electrostatic
calculation of the lens strength of a superconducting film with a hole
follows in Sec. \ref{sec.lens}. Section \ref{sec.disc} concludes with
a discussion.

\section{Boundary conditions for {\lowercase{\it d}}-wave order parameter}
\label{sec.bound}

A spatially varying order parameter profile can be described in the
framework of the phenomenological Ginzburg-Landau (GL) theory. The
physics near the surface crucially depends on the imposed boundary
conditions. As shown by Gor'kov, this phenomenological theory can be
derived from the microscopic Bardeen-Cooper-Schrieffer theory
\cite{dG66}. We essentially follow this approach in order to determine
the boundary conditions appropriate for HTSC.

The fact that the tetragonal HTSC materials ${\rm YBa}_2{\rm Cu}_3
{\rm O}_{7-\delta}$ and ${\rm La}_{2-x}{\rm Sr}_x{\rm CuO}_4$ belong
to the class of unconventional $d$-wave superconductors appears has
been reliably established \cite{d-wave}. The term ``unconventional''
means that the spatial symmetry of the superconducting order parameter
$\Delta_{{\bf k}, \alpha\beta} \sim \langle a_{{\bf k}\alpha} a_{-{\bf
    k}\beta} \rangle$ is lower than in the normal state \cite{SU91}.

More specifically evidence favors a $d_{x^2-y^2}$-symmetry. In this
case the order parameter can be written in the form
\begin{equation}
  \Delta_{{\bf k},\alpha\beta}({\bf r}) = (i\sigma_y)_{\alpha\beta}
  \psi({\bf k}) \Delta({\bf r})
\end{equation}
with the Pauli matrix $\sigma_y$. The momentum dependence at the Fermi
surface is described by the following normalized basis function of the
irreducible representation $B_{1g}$ of the tetragonal group $D_{4h}$:
\begin{equation}
\label{def.psi}
 \psi({\bf k})=\frac{\sqrt{15}}{2}(\hat k_X^2-\hat k_Y^2),
\end{equation}
where $X$ and $Y$ are internal axes of the crystal, and $\hat{\bf k}=
{\bf k}/|{\bf k}|$.

In the vicinity of $T_{\rm c}$ the gap profile $\Delta({\bf r})$ near
the superconductor/insulator interface is determined by the solution
of the GL equation
\begin{eqnarray}
\label{GL.eqs}
 -\xi_{\parallel}^2(T) \left(\frac{\partial^2\Delta}{\partial X^2}+
   \frac{\partial^2\Delta}{\partial Y^2}\right) & - &
  \xi_{\perp}^2(T)\frac{\partial^2\Delta}{\partial Z^2}
    \nonumber \\
   & &-\Delta+\frac{1}{|\Delta_0|^2} |\Delta|^2\Delta=0.
\end{eqnarray}
Explicit values of the correlation lengths $\xi_{\parallel(\perp)}(T)$
along different axes and the gap saturation amplitude $\Delta_0$ are
given in Appendices [Eqs. (\ref{gapcoeff}) and (\ref{xis})].

\begin{figure}
\begin{center}
\leavevmode
\epsfxsize=0.5 \linewidth
\epsfbox{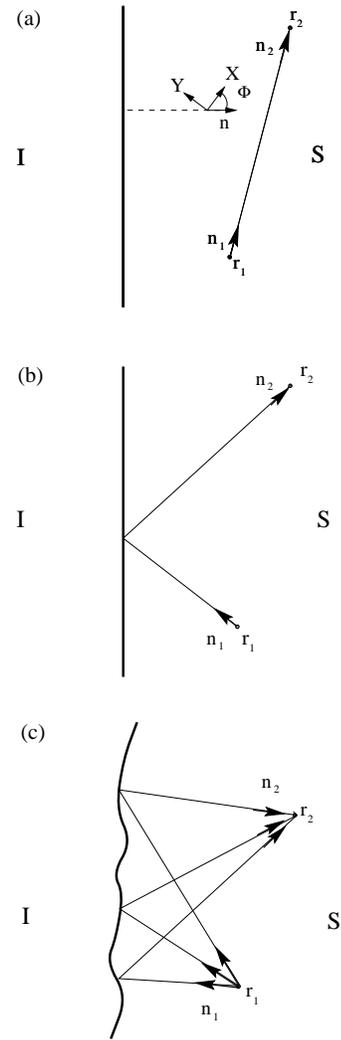}
\caption{Electron paths of the quasiclassical approximation: 
  (a) straight paths contribute to the kernel independent of the
  surface, (b) for specular reflecting surfaces one additional
  trajectory contributes (c) for diffusively reflecting surfaces a
  whole group of additional trajectories contributes.}
\label{fig.paths}
\end{center}
\end{figure}

To obtain the boundary condition for the GL equation (\ref{GL.eqs})
microscopically, we start by noting that the Gor'kov equations take
the form of a linearized integral equation near $T_{\rm c}$
\cite{dG66}:
\begin{equation}
\label{int.eqn}
 \Delta({\bf r}_1)=\int d^3r_2~K({\bf r}_1,{\bf r}_2)\Delta({\bf r}_2).
\end{equation}
The kernel $K$ can be calculated in the quasiclassical approximation,
making use of the method of classical trajectories
\cite{dG66,Luders}. Depending on the roughness of the surface,
different kinds of the reflection of electrons, usually referred to as
diffusive and specular, are to be considered (see Fig.~\ref{fig.paths}).

We consider the following geometry: The normal vector ${\bf n}$ to the
superconductor/insulator interface lies in the basis $XY$ plane, the
angle between ${\bf n}$ and the axis $OX$ of the underlying tetragonal
lattice being equal to $\Phi$. The interface is assumed to be
macroscopically flat, i.e. the roughness is restricted to scales
smaller than the correlation length. We denote $\xi
\equiv \xi_\parallel$ unless stated otherwise.
 
Let us start with the case of diffusive reflection for microscopically
rough surfaces. In the absence of an external magnetic field the order
parameter depends only on the distance $x$ from the surface. It is
convenient to introduce the dimensionless coordinate $\tilde
x=x/\xi_0$ ($\xi_0= {v_{\rm F}}/{2\pi T_{\rm c}}$ is the coherence
length), and the gap equation (\ref{int.eqn}) takes the form
\begin{equation}
\label{gap.eqn}
 \Delta(\tilde x_1)=\int_0^{\infty} d\tilde x_2~K(\tilde x_1,\tilde x_2)
      \Delta(\tilde x_2),
\end{equation}
where
\beginwidetop
\begin{eqnarray}
\label{kernel}
K(\tilde x_1,\tilde x_2)=\frac{VN_0}{2} \sum\limits_n & \Biggl[ &
    \int_0^1\frac{ds}{s}~F_0(s)
     \exp\left(-\frac{|2n+1|}{s}|\tilde x_1-\tilde x_2|\right) \nonumber \\
    &+& \int_0^1ds_1\int_0^1ds_2~F_{\rm r}(s_1,s_2)
    \exp\left(-|2n+1|(\frac{\tilde x_1}{s_1}+\frac{\tilde x_2}{s_2})\right)
    \Biggr],
\end{eqnarray}
\endwide where $V$ is the coupling constant, and $N_0$ is the electron
density of states. The kernel $K$ in (\ref{kernel}) is composed of a
bulk and a surface contribution. Both crucially depend on the angle
$\Phi$ through the functions $F_0$ and $F_r$:
\begin{eqnarray}
\label{F0.dwave}
 F_0(s)=\frac{15}{32}(3-14s^2+19s^4)\cos^2 2\Phi \nonumber \\
   +\frac{15}{2}(s^2-s^4)\sin^2 2\Phi,
\end{eqnarray}
\begin{equation}
\label{Fdiff.dwave}
  F_{\rm r}(s_1,s_2)=\frac{15}{8}(1-3s_1^2)(1-3s_2^2)\cos^2 2\Phi.
\end{equation}
Details of the derivation of these expressions can be found in
Appendix~\ref{app.quasi}.

In order to obtain the boundary condition we have to evaluate the
linearized gap equation (\ref{gap.eqn}) with the kernel
(\ref{kernel}). We will do this at the critical temperature $T_{\rm
  c}$, where the BCS condition
\begin{equation}
  \int_{-\infty}^{+\infty}d\tilde x_2~K_0(\tilde x_1-\tilde x_2)=1,
\end{equation}
holds. Here $K_0$ is the bulk part of (\ref{kernel}). 
In the region $1 \ll \tilde x\ll \xi(T)/\xi_0$ the surface
contribution of the kernel is small compared to the bulk contribution
and the non-linearity of the gap equation still can be neglected.
Therefore the gap equation is solved by a linear function,
$\Delta(\tilde x)=\Delta(0)(1+\tilde x/\tilde b)$, so that the
effective boundary condition for the order parameter in the GL region
takes the form
\begin{equation}
\label{bc.x}
 \left. \frac{\partial \Delta}{\partial\tilde x}\right|_{\tilde x=0}
        =\left. \frac{1}{\tilde b} \Delta\right|_{\tilde x=0}.
\end{equation}
The ``extrapolation length'' $b$ acquires from the kernel a dependence
on the orientation of the surface with respect to the underlying
crystal lattice.

We evaluate the parameter $b$ using the variational approach of Ref.
\cite{Sam1}. It is convenient to introduce the function
$q(\tilde x)$ by $\Delta(\tilde x)=C(\tilde x+q(\tilde x))$, from
which the parameter $\tilde b$ follows according to $\tilde
b=\lim\limits_{\tilde x\to\infty} q(\tilde x)$. The gap equation can
now be rewritten as
\begin{equation}
\label{eqq}
q(\tilde x_1)=\frac{1}{2} E(\tilde x_1)+\int_0^{\infty }d\tilde x_2~
   K(\tilde x_1,\tilde x_2)q(\tilde x_2),
\end{equation}
where
\beginwide
\begin{eqnarray*}
 E(\tilde x_1)&=&2\int_0^{\infty}d\tilde x_2~\tilde x_2
  K(\tilde x_1,\tilde x_2)-2\tilde x_1 \\
 &=&VN_0 \sum\limits_n\frac{1}{(2n+1)^2} \Biggl[ \int_0^1 ds~s
 F_0(s)\exp\left(-\frac{|2n+1|}{s}\tilde x_1\right)
 +\int_0^1ds_1\int_0^1ds_2~s_2^2F_{\rm r}(s_1,s_2)
 \exp\left(-\frac{|2n+1|}{s_1}\tilde x_1\right)\Biggr].
\end{eqnarray*}
\endwide Apart from a prefactor, the solution of Eq. (\ref{eqq}) can
be obtained by minimizing the functional
\begin{equation}
\label{func}
 {\cal F}[q]=\frac{\int_0^{\infty }d\tilde x~q(\tilde x)\left[
   q(\tilde x)-\int_0^{\infty }d\tilde x^{\prime }
    K(\tilde x,\tilde x^{\prime })q(\tilde x^{\prime })\right]}
    {\left[\int_0^{\infty }d\tilde x~q(\tilde x)E(\tilde x)\right]^2}
\end{equation}
with respect to $q$. The minimum value is given by
\begin{equation}
\label{fmin}
 {\cal F}_{\rm min}=\frac{1}{2\int_0^{\infty }d\tilde x~E(\tilde x)
    q(\tilde x)}.
\end{equation}

The extrapolation length can be related to this minimum value as
follows: Eq. (\ref{eqq}) can be rewritten as an equation for $Q(\tilde
x)=q(\tilde x)-\tilde b$, which vanishes for $\tilde x\to\infty$. Upon
multiplying both sides of the equation for $Q$ by $\tilde x_1$,
integrating over $\tilde x_1$ and taking into account (\ref{fmin}), we
finally obtain \cite{Sam2} (see also \cite{Agter}) an exact expression
for the extrapolation length $b=\tilde b\xi_0$:
\begin{eqnarray}
\label{bexact}
\frac{b}{\xi_0} = \frac{1}{7\zeta(3)}\frac{1}{\int_0^1 ds~s^2F_0(s)}
     \Biggl\{ \frac{\pi^4}{24}\Bigl[\int_0^1 ds~s^3F_0(s) \nonumber \\
 +\int_0^1 ds_1\int_0^1 ds_2~s_1^2s_2^2F_r(s_1,s_2)\Bigl]+
  \frac{1}{VN_0{\cal F}_{\rm min}} \Biggr\},
\end{eqnarray}
where $\zeta(x)$ is the Riemann zeta-function.

Now we are able to apply the variational principle. Substituting in 
(\ref{func}) unity as a trial function, we have:
\beginwidetop
\begin{equation}
\label{b}
  VN_0{\cal F}_{\rm min}=\frac{2\pi^2}{49\zeta^2(3)}
     \frac{\int_0^1 ds~sF_0(s)-\int_0^1 ds_1
         \int_0^1 ds_2~s_1s_2F_r(s_1,s_2)}{\Bigl[
     \int_0^1 ds~s^2F_0(s)+\int_0^1 ds_1
         \int_0^1 ds_2~s_1s_2^2F_r(s_1,s_2)\Bigr]^2}.
\end{equation}
\endwide

As one can see from Appendix \ref{app.quasi}, the corresponding
expressions for the specular case follow from Eqs. (\ref{bexact}) and
(\ref{b}) by replacing $F_{\rm r}(s_1,s_2)$ with
$\frac{1}{s_2}\delta(s_1-s_2)F_{\rm r}(s)$, where $F_{\rm r}(s)$ is
given by
\begin{eqnarray}
\label{Fspec.dwave}
 F_{\rm r}(s)=\frac{15}{32}(3-14s^2+19s^4)\cos^22\Phi \nonumber \\
   -\frac{15}{2}(s^2-s^4)\sin^22\Phi.
\end{eqnarray}
Then the results of Ref.\cite{Sam1} are recovered.

\begin{figure}
  \epsfxsize=0.9 \linewidth 
\epsfbox{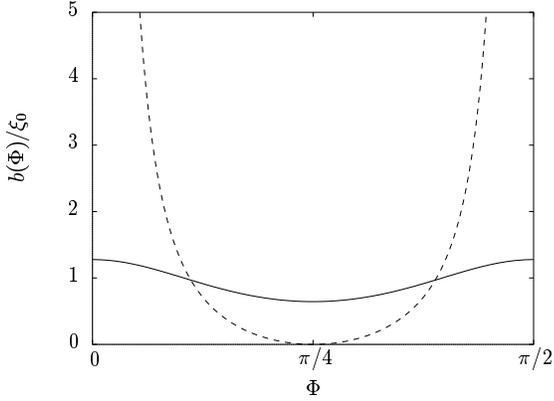}
\caption{Extrapolation length $b$ as a function of the angle 
  $\Phi$ for the diffusive (solid line) and the specular (dashed line)
  case.}
\label{fig.b}
\end{figure}

The extrapolation lengths for the diffusive and specular cases, which
result from the substitution of expressions (\ref{F0.dwave}),
(\ref{Fspec.dwave}), and (\ref{Fdiff.dwave}) in (\ref{b}), are plotted
in Fig. \ref{fig.b} as a function of the angle $\Phi$.

The profile of the order parameter near the surface is determined by
the ratio of the extrapolation length $b$ and the correlation length
$\xi(T)$. In the diffusive case the value of the
extrapolation length $b$ just slightly oscillates as a function of
$\Phi$. For all orientations of the surface it is of the order of the
coherence length $\xi_0$. In the GL region it is therefore much
smaller than the characteristic scale of the order parameter, the
correlation length $\xi(T)$, and the boundary condition for the GL
equation effectively becomes
\begin{equation}
\label{bc.diff}
 \Delta|_{x=0}=0.
\end{equation}

In the specular case the ratio $b(\Phi)/\xi_0$ strongly oscillates
from $0$ to $\infty$. However, in the GL region $b(\Phi) \ll \xi(T)$
for most orientations, and effectively $\Delta|_{x=0}=0$ again. Only
in a narrow range of orientations $b\gg \xi$, where the boundary
condition is effectively
\begin{equation}
\label{bc.spec}
 \left. \frac{d\Delta}{dx}\right|_{x=0}=0.
\end{equation}

In the consideration given above, the normal vector ${\bf n}$ was assumed
to lie in the basis plane of a tetragonal crystal. If ${\bf n}$ is
directed along the fourth order axis, then the functions $F_0$ and
$F_{\rm r}$ take the following form:
\begin{equation}
\label{Fspec.z}
 F_0(s)=F_{\rm r}(s)=\frac{15}{8}(1-s^2)^2,
\end{equation}
\begin{equation}
\label{Fdiff.z}
 F_{\rm r}(s_1,s_2)=0
\end{equation}
[see (\ref{F0}), (\ref{Fspec}) and (\ref{Fdiff})]. As follows from
(\ref{Fspec.z}), the {\it exact} boundary condition along
$z$-direction for specular reflection is (cf. Refs.\cite{SU91,Sam1})
\begin{equation}
\label{bc.spec.z}
 \left. \frac{d\Delta}{dz}\right|_{z=0}=0. 
\end{equation}
In the diffusive case the corresponding gap equation can, in
principle, be solved exactly since it has a Wiener-Hopf form due to
(\ref{Fdiff.z}). However, we shall not proceed in this rather
cumbersome way but instead use the variational method again, which
gives $b\approx 0.46\xi_0$.

Thus, the final conclusion is the following: We are able to use the
boundary condition (\ref{bc.diff}) for {\it all} orientations of a
diffusively reflecting surface with a normal vector lying in the $XY$
plane. For a specularly reflecting surface this condition holds for
{\it almost all} orientations, except from very narrow angular regions
near $\Phi=0,\frac{\pi}{2}$ {\it etc}, where we should use
(\ref{bc.spec}) instead.  The width of these regions is of the order
of ${\xi_0}/{\xi(T)}$, which is negligibly small near $T_{\rm c}$.

The profile of the order parameter at the surface is determined by the
solution of the GL equation (\ref{GL.eqs}) subject to the boundary
condition (\ref{bc.x}). One obtains
\begin{equation}
\label{profile.OP}
\Delta(x)=\Delta_0 \tanh[(x-x_0)/\sqrt2 \xi(T)]
\end{equation}
with
\begin{equation}
x_0=-\sqrt{2}\xi(T){\rm arctanh}\left(-\frac{\xi(T)}{\sqrt{2}b}+\sqrt{1+
\frac{\xi^2(T)}{2b^2}}\right).
\end{equation}
In the case of a diffusively reflecting boundary we have obtained $b
\sim\xi_0$. Thus, in this case it is reasonable to use the estimate
$x_0 \approx 0$ near $T_{\rm c}$.

\section{Surface charging}
\label{sec.surf}

To obtain the charge modulation near an infinite plane surface we
follow the calculations for the charging of a flux line by Blatter {\em
et al.} \cite{B+96}. The spatial variation in the order parameter
induces a modulation of the chemical potential $\mu$. As derived in
Appendix \ref{app.dwp} [Eq. (\ref{del.n})], this generates a variation of
the local charge density
\begin{eqnarray}
  \varrho_{\rm ext}(x)&=&-e N'_0 \ln\frac{\epsilon_{\rm c}}{T_{\rm c}}
  \left[\Delta_0^2-\Delta^2(x)\right]\nonumber \\
  &\approx& -4e N'_0
  \Delta_0^2 \ln\left(\frac{\epsilon_{\rm c}}{T_{\rm c}}\right)
 \exp\left(-\frac{\sqrt{2}x}{\xi}\right)
\end{eqnarray}
where $N_0'$ is derivative of the density of states at the Fermi
level, $\epsilon_{\rm c}$ is the energy cutoff of the electron
intraction, $\Delta_0$ denotes the bulk value of the gap function,
and $N_0$ denotes the density of states at the Fermi level. We
consider $\varrho_{\rm ext}(x)$ as ``external'' charge density since
screening has not yet been taken into account.

Metallic screening can be included within a Thomas-Fermi approximation
by introducing the spatially constant electrochemical potential
$\mu+e\phi$ instead of $\mu$. The electrostatic potential $\phi$ is
then determined in linear order by the one-dimensional screened
Poisson equation:
\begin{equation}
\left[\frac{d^2}{dx^2}-\frac{1}{\lambda_{\rm TF}^2}\right] \phi(x) =
4\pi \varrho_{\rm ext}(x).
\end{equation}
The Thomas-Fermi screening length is given by $\lambda_{\rm TF}=(8\pi
e^2 N_0)^{-1/2}$.

Overall charge neutrality in the superconducting halfspace $x>0$
requires a solution that meets the boundary condition $\phi'(x=0)=0$.
In the limit $\lambda_{\rm TF} \ll \xi$ this requirement is fulfilled
by the solution
\begin{eqnarray}
\label{sol-phi}
  \phi(x)&\approx&-16\pi e N'_0 \Delta_0^2 \frac{\lambda_{\rm TF}^2}{\xi}
  \ln\left(\frac{\epsilon_{\rm c}}{T_{\rm c}}\right)\nonumber \\ 
  &&\times \left[\sqrt{2}\lambda_{\rm
      TF}e^{-x/\lambda_{\rm TF}}-\xi e^{-\sqrt{2}x/\xi}\right].
\end{eqnarray}
The corresponding screened charge density is
$\varrho(x)=-\phi''(x)/4\pi$. In the case of $d$-wave pairing one gets,
with the specific gap (\ref{gapcoeff}) and correlation length
(\ref{xis}) for a three-dimensional parabolic band, the explicit
expression
\begin{eqnarray}
\label{chdensity}
\varrho(x)&\approx& \frac{1}{5} \frac{ea_{\rm B}}{\xi^3}
\ln\left(\frac{\epsilon_{\rm c}}{T_{\rm c}}\right)
\nonumber \\
&&\times \left[\frac{\sqrt{2}}{\lambda_{\rm TF}}e^{-x/\lambda_{\rm TF}}
  -\frac{2}{\xi}e^{-\sqrt{2}x/\xi}\right]
\end{eqnarray}
with the Bohr atom radius $a_{\rm B}$.  

\begin{figure}
\begin{center}
\leavevmode
\epsfxsize=0.9 \linewidth
\epsfbox{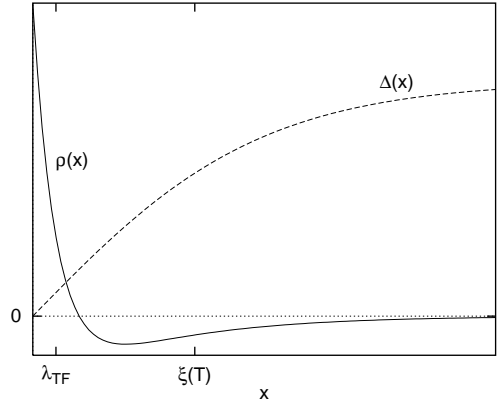}
\caption{Modulation of the gap $\Delta(x)$ and the charge density
  $\rho(x)$ at an infinite plane surface as a
  function of the normal distance $x$ from the surface (arbitrary units).}
\label{fig.profile}
\end{center}
\end{figure}

Due to screening this charge distribution forms a double layer of
opposite charge, see Fig. \ref{fig.profile}. The outer layer with a
thickness of the order of $\lambda_{\rm TF}\ln(\xi/\sqrt{2}
\lambda_{\rm TF})$ contributes a total charge
\begin{equation}
Q\approx\frac{\sqrt{2}}{5}\frac{ea_{\rm B}}{\xi^3}
\end{equation}
per unit surface area. Here we have used that $\ln(\epsilon_{\rm c}
/{T_{\rm c}})$ is usually of order of unity for high--$T_{\rm c}$
superconductors. The corresponding dipole moment $g$ per unit area of
the surface is given by
\begin{equation}
\label{dipolemoment}
g=\int_0^{\infty} x \varrho(x) dx \approx -\frac{\xi}{\sqrt{2}}Q.
\end{equation}

\section{Lens effect}
\label{sec.lens}

\begin{figure}
\begin{center}
\leavevmode
\epsfxsize=0.7 \linewidth
\epsfbox{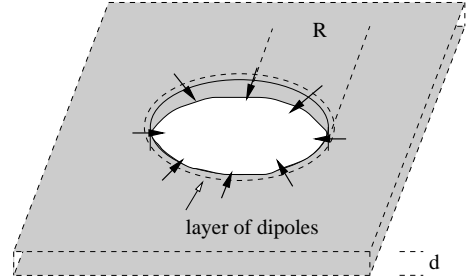}
\caption{Superconducting film with circular hole inside acting as
  lens for a penetrating electron beam.}
\label{fig.lens}
\end{center}
\end{figure}

So far we have derived that the surface of a high--$T_{\rm c}$
superconductor can induce a charge modulation. Now we comment on a
possible experimental observation of this effect. One setup, an
``electrooptical lens'', is shown in Fig.~\ref{fig.lens} and can be
considered as idealization of a recent experiment \cite{Kri95}.
Motivated by the observed lens effect, which sets in when the material
becomes superconducting, we will examine to what extent it can be
ascribed to a surface charging according to the presented mechanism.

Let us assume that the sample is a thin $d$-wave superconducting
monocrystalline film with a micro-size circular hole inside. The
surfaces of the film are assumed to be smooth and parallel to the $XY$
crystal plane. Assuming specular reflection, the order parameter is
not suppressed at these surfaces according to Eq.  (\ref{bc.spec.z}).
But the inner surface of the hole is normal to the $XY$ plane leading
to a strongly suppressed gap function at the boundary of the hole. Due
to the preparation of the hole this surface should be rough leading
to a decrease of the order parameter around the hole according to Eq.
(\ref{bc.diff}). Thus one obtains a charge accumulation around the
hole only \cite{note}. Since the real charge distribution consists of
two layers of positive and negative charge one can think of this
distribution as a ring of dipoles oriented normal to the boundary of
the hole, see Fig.~\ref{fig.lens}. The corresponding electrostatic
field should be observable in different ways. One possible way to
detect the ring of dipoles is to send a very slow and weak beam of
electrons through the hole. The intensity in the center of the beam
image behind the hole then changes when the sample is cooled below
$T_{\rm c}$ due to the (de)focusing caused by the accumulated dipoles.

In the actual experiment by Kriebel {\it et al.} \cite{Kri95} a similar
geometry was examined by transmission electron microscopy. They have
observed a very small change in the image intensity of the electron
beam at the superconducting transition. However, the actual sample was
polycrystalline and can not be expected to lead to such a well-defined
field distribution as in our ideal situation described above.
Nevertheless, to obtain the order of magnitude of this lens effect, it
is sufficient to consider the simplified setup.

The calculation of the lens effect amounts to solving the
electrostatic potential problem for the ring of dipoles. From the
electric field we calculate the optical effect in an eikonal
approximation for the electrons.

Since we can assume that the radius of the hole is much larger than
the correlation length $\xi$, the curvature of the boundary can be
neglected.  Therefore the accumulated charge can be approximated by
the charge density (\ref{chdensity}) obtained for a plane surface.
With the (constant) dipole moment $g$ per unit area from Eq.
(\ref{dipolemoment}) the electric potential outside the
superconducting film is given by
\begin{equation}
  \phi({\bf r})=\frac{g}{4\pi}\int_{\rm ring} {\bf n}\cdot {\bbox \nabla}
  \left(\frac{1}{|{\bf r}-{\bf r'}|}\right) dS'
\end{equation}
with the surface normal vector ${\bf n}$ lying in the $XY$ plane and
the surface element $dS'$ of the ring. For a cylinder with radius
$R$ and height $d \ll R$ this leads to
\begin{equation}
\label{pot}
  \phi(\rho,z)=\frac{gd}{4\pi}\int_0^{2\pi} d\varphi \frac{R\rho
  \cos\varphi - R^2}{\left[\rho^2+z^2+R^2-2R\rho\cos\varphi
  \right]^{3/2}}.
\end{equation}
This integral can be expressed in terms of complete elliptic
integrals. But we are only interested in the radial electric field
component averaged along the $z$ direction to obtain the total
intensity change. Let us assume that an electron moves initially at a
radial distance $\rho_i$ from the center in the negative $z$-direction
with vanishing radial velocity. To minimize the influence of the
source of the electron on the effect one should assume that the source
is asymptotically far away.

\begin{figure}
\epsfxsize=0.9 \linewidth
\epsfbox{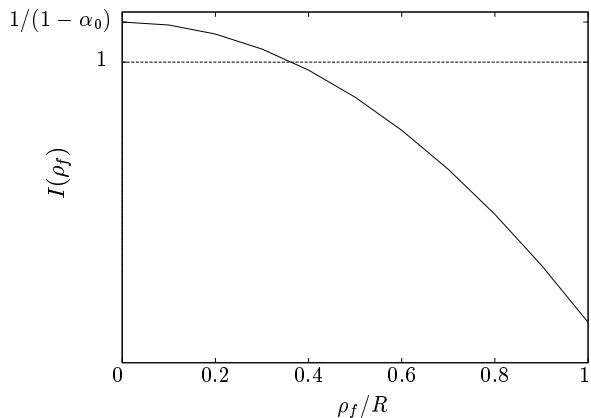}
\caption{Variation of the relative intensity $I(\rho_f)$ of the
  electron beam image across the hole behind the superconducting
  film.}
\label{fig.contrast}
\end{figure}

For this radial distance $\rho_i$ from the $z$-axis the averaged value
of the deflecting field for $-L<z<0$, defined by
\begin{equation}
\overline{E_{\rho}}(\rho_i)=\frac{1}{L}\int_{-L}^0 dz E_\rho(\rho_i,z),
\end{equation}
can be calculated as follows: Consider a virtual cylinder of radius
$\rho_i$ around the $z$-axis with its ends at $z=0$ and $z=-L$,
respectively.  For symmetry reasons the flux of the electric field
across the abutting face at $z=0$ vanishes. If the lower face of the
cylinder is far away from the dipoles, i.e.  $L\gg R$, the electric
potential at the lower face can be approximated by an asymptotical
expansion of (\ref{pot}) in $z$.  Then the angular integration is
trivial and one obtains for the $z$-component of the asymptotic
electric field
\begin{equation}
\label{asympt}
  E_z(\rho,z)=\frac{3gdR^2}{4}\frac{2z^2-10\rho^2-5R^2}{z^6}+
  {\mathcal{O}}(z^{-8}).
\end{equation}
Because there is no charge located inside the virtual cylinder, the
total flux of the electric field across the surface of the cylinder
vanishes. Using the asymptotic value (\ref{asympt}) to calculate the flux
across the face at $z=-L$ one obtains the averaged radial field for $z<0$
\begin{equation}
\label{aver}
\overline{E_{\rho}}(\rho_i)\approx \frac{3}{8}g d \rho_i R^2
\frac{5(\rho_i^2+R^2)-2L^2}{L^7}.
\end{equation}
For symmetry reasons the same expression holds for an average of the
electric field in the upper half space. We take $L\rightarrow \infty$
as starting position of the electrons. Therefore the electric field in
the upper half space does not contribute to the total deflection of
the electron, as can be seen easily from (\ref{aver}).  Since the
whole change in the intensity comes then from the path of the electron
behind the film, the optimal position $z=-L$ of the detector has to be
obtained by a maximization of the averaged radial field (\ref{aver})
as a function of $L$. For electrons near by the center of the hole
($\rho_i\approx 0$) the optimal distance is $L=(\sqrt{14}/2) R$.  For
this rather small value of $L$ it can be checked numerically that the
asymptotic approximation used for the average procedure underestimates
$\overline{E_{\rho}}$ only by a factor $\approx 2$ if $\rho_i \ll R$.
Using this optimized value, the electron will be detected at $z=-L$
with a radial distance $\rho_f(\rho_i)$ from the center given by
\begin{equation}
\label{intens}
\rho_f(\rho_i)=\rho_i(1-\alpha_0+\alpha_2 \rho_i^2)
\end{equation}
with
\begin{eqnarray}
\label{param}
\alpha_0 \approx 0.016 \frac{e}{m} \frac{gd}{R v_0^2}, \\
\alpha_2 \approx 0.04 \frac{e}{m} \frac{gd}{R^3 v_0^2},
\end{eqnarray}
where $v_0$ is averaged velocity of the electron with respect to the
$z$ direction. The relative enhancement of the intensity at the center
is then given by
\begin{equation}
\left. \frac{d\rho_i}{d\rho_f}\right|_{\rho_f=0}=\frac{1}{1-\alpha_0}.
\end{equation}
Using the density of the dipole moment (\ref{dipolemoment}) we obtain
for the deflection parameter
\begin{equation}
\label{defl}
\alpha_0 \approx -3.2 \cdot 10^{-3} \frac{e^2 a_{\rm B}}{m v_0^2 \xi^2}
\frac{d}{R} \approx -\frac{10^{-4}{\rm eV}}{E_{\rm kin}}\frac{d}{R},
\end{equation}
where we have used $\xi \approx 10${\AA} as a typical value for
high--$T_{\rm c}$ superconductors in the last expression. The spatial
variation of the relative intensity $I(\rho_f)=d\rho_i/d\rho_f$
determined by (\ref{intens}) is shown in Fig. \ref{fig.contrast}.
Strictly, the shown numerical result is valid only for distances
$\rho_f\ll R$ due to the truncation of higher order terms of $\rho_i$
in Eq. (\ref{intens}).

\section{Discussion}
\label{sec.disc}

In Sec. \ref{sec.bound} we have presented an explicit microscopic
calculation of the extrapolation length in $d$-wave superconductors.
Within the quasiclassical approximation we found $b \sim \xi_0$, which
is of the order of 12{\AA} in YBCO. Within the same approximation
microscopic calculations\cite{dG64} for $s$-wave superconductors give
$b=\infty$, i.e. no suppression of the order parameter at the surface.
In the $s$-wave case a finite value of 
\begin{equation}
\label{b.dG}
b \sim \xi_0^2/a
\end{equation}
can be found only if one goes {\em beyond} the quasiclassical
approximation, taking special care of the finite thickness of the
metal-vacuum transition layer\cite{CGM62,dG64}. From this expression
one can expect small $b$, i.e. a notable suppression of the condensate
near the surface, if $\xi_0$ is small (i.e. of the order of $a$). This
is the case for HTSC as has been pointed out by Deutscher and
M\"uller\cite{DM87}. However, Eq. (\ref{b.dG}) is based on
approximations which are valid only as long as $b \gg \xi_0$. For
$d$-wave superconductivity the situation is very different: as our
calculation shows, the most important reason for the reduction of $b$
is the $d$-wave symmetry itself. In this case the effect is so
pronounced that it can be obtained within the quasiclassical
approximation.

It should be emphasized that in this article we assume the
superconducting order parameter to have the pure
$d_{x^2-y^2}$-symmetry [see Eq. (\ref{def.psi})], which immediately
follows from the chosen form of the pairing potential (\ref{spec.V})
containing the spherical harmonics corresponding only to the
irreducible representation $B_{1g}$ of the tetragonal group. In
general, one could allow for the possibility of a small admixture of
the $s$-wave component. As shown in Ref. \cite{s+d}, such an admixture
appears whenever one deals with a spatially non-uniform distribution
of the order parameter, e.g. near surfaces, twinning planes, columnar
defects {\it etc.} and is brought about by the presence of the
mixed-gradient terms in the GL expansion of the free energy. In our
work we neglect this effect due to the results of Ref. \cite{s+d}: (i)
The induced $s$-wave component is always much smaller (by one order of
magnitude at least) than the $d$-wave one. This holds in particular at
temperatures close to $T_{\rm c}$, since the $s$-wave component
vanishes faster with $(T_{\rm c}-T)$ than the $d$-wave component.
(ii) The $s$-wave component {\it vanishes} at the surface together
with the $d$-wave component, so that the modulation of $\rho(x)$ shown
in Fig.  \ref{fig.profile} remains unaffected.

The fact, that the $d$-wave order parameter is strongly suppressed at
diffusively reflecting superconductor/insulator interfaces, could
explain an anomalously weak temperature dependence of the Josephson
critical current $j_{\rm c}\sim (T_{\rm c}-T)^2$ near $T_{\rm c}$, as
observed in Ref. \cite{Gross90}. Indeed, the critical current is
proportional to the square of the order parameter at the surface:
$j_{\rm c}\sim\Delta^2|_{\rm S}$. However, as it follows from our
results [see Eq. (\ref{profile.OP})]: $\Delta|_{\rm S}\sim
(b/\xi(T))\Delta_0 \sim(\xi_0/\xi(T))\Delta_0$. Hence
\begin{equation}
j_{\rm c}\sim\Delta_0^2(T)\frac{\xi_0^2}{\xi^2(T)} \sim(T_{\rm c}-T)^2
\end{equation}
for all orientations, instead of $j_{\rm c}\sim T_{\rm c}-T$, as
expected for a conventional SIS junction.  The same result has been
obtained in \cite{BGZ95} for distinguished orientations of a
specularly reflecting interface (see also the discussion in the end of
Sec. \ref{sec.bound}).

In the present analysis we have neglected possible anisotropy of the
Fermi surface, which is characteristic for HTSC. In terms of the ratio
of effective masses in a Ginzburg-Landau theory this anisotropy can be
of the order $m_\perp /m_\parallel \sim 100$. The generalization of
our analysis to anisotropic Fermi surfaces is straightforward but
rapidly leads to cumbersome analytic expressions. Such a
generalization has been performed by Shapoval \cite{Sha85} for the
case of an anisotropic $s$-wave pairing. For $d$-wave pairing we
expect that this generalization leads only to quantitative changes
which only weakly affect the boundary condition $b \ll \xi(T)$.
Although $N_0'$ and $\lambda_{\rm TF} \sim N_0^{-1/2}$ drastically
depend on the mass anisotropy. In contrast the combination $N_0'
\lambda_{\rm TF}^2$ is insensitive. Therefore only the spatial
distribution of charge $\varrho$ will be affected, but not the total
charge $Q$ or the dipole moment $g$.

The amplitude of the electrooptical lens effect, Eq. (\ref{defl}),
crucially depends on the beam energy of the electrons. In the
experiment of Ref. \cite{Kri95} a beam energy of $E_{\rm kin} =
120$keV was used, which leads to $\alpha_0 \sim 10^{-9}$ at the
optimum distance $L \sim R$ behind the lens. However, the actual
picture was generated much further away, where the intensity rapidly
decreases according to Eq. (\ref{aver}). In addition, the sample had a
much more irregular shape than in our idealized geometry. We expect
this irregularity to reduce the electrooptical effect of the dipole
layer. Therefore we believe that surface charging has to be excluded
as the origin of the lens effect with $\alpha_0 \sim 10^{-2}$ observed
in this particular experiment.

At present, the actual origin for the observed effect can not be
deduced unambiguously from the experiment. One contribution could be
due to a charge accumulation at the surface of the superconductor
which is directly hit by the electron beam. The amount of accumulated
charge then strongly depends on the conductivity of the sample which
drastically changes at $T_{\rm c}$. Since in the experiment
polycrystalline samples have been used, there are small magnetic
fields induced by the Josephson currents between the grains. But these
fields should give an even weaker deflection of the electrons since
the local field sources do not sum up constructively.

In order to clarify this situation it would be desirable to use lower
electron energies where the the electrooptical effect should become
stronger. For $E_{\rm kin} = 100$eV as used in LEED, $\alpha_0 \sim
10^{-6}$ can be achieved due to surface charging at the optimum
distance. In addition, the use of a single crystalline sample with a
more regular shape of the hole would be desirable as well as a
spatial resolution of the density profile.

Finally, to detect the charging effect, other experimental techniques
might be more promising than the lens effect examined above. For
example, the local stray field outside the superconductor may be
probed by atomic force microscopy, as already suggested by Blatter
{\em et al.} \cite{B+96} for vortices.

\section*{Acknowledgments}

The authors gratefully acknowledge helpful discussions with B.
B\"uchner, R. Gross, and O. Hoffels and a critical reading of the
manuscript by L.K. Sundman. This work was supported by the Deutsche
Forschungsgemeinschaft SFB 341 and by the German--Israeli Foundation
(GIF).

\appendix

\section{{\lowercase{\it d}}-wave pairing}
\label{app.dwp}

We briefly point out the most important changes for
$d_{x^2-y^2}$-pairing in contrast to conventional $s$-wave pairing in
a weak-coupling BCS model. A discussion of arbitrary symmetry has been
summarized in Ref. \CITE{SU91}. For calculational convenience we
ignore mass anisotropy and use a dispersion relation $\epsilon_{\bf k}
= k^2/2m - \mu$ of single electron states. They have a density of
states (per spin projection)
\begin{equation}
  N_\epsilon:=\int \frac{d^3k}{(2 \pi)^3}
  \delta(\epsilon-\epsilon_{\bf k}) = \frac{[2 m^3
    (\epsilon+\mu)]^{1/2}}{2 \pi^2}.
\end{equation}

Irrespective of specific pairing mechanism, the gap function for
singlet Cooper pairs with relative momentum ${\bf k}$ can be written
as
\begin{equation}
\label{def.gap}
\Delta_{{\bf k},\alpha\beta}= 
(i \sigma_y)_{\alpha \beta} \psi({\bf k}) \Delta_0 
=\int \frac{d^3k'}{(2 \pi)^3} V({\bf k},{\bf k}') \langle a_{{\bf k}' 
    \alpha} a_{-{\bf k}' \beta}\rangle.
\end{equation}

For a singlet state the potential can be approximated by
\begin{equation}
\label{spec.V}
V({\bf k}, {\bf k}')=
\left\{
\begin{array}{ll}
-V \psi({\bf k}) \psi^*({\bf k}'), &
|\epsilon_{\bf k}|, |\epsilon_{{\bf k}'}| \leq \epsilon_{\rm c}\\
0, & {\rm otherwise}
\end{array}
\right.
\end{equation}
with an energy cutoff $\epsilon_{\rm c}$ of the interaction. For
anisotropic masses the definition (\ref{def.psi}) of $\psi$ was
normalized such that $\langle \psi^2({\bf k}) \rangle_{0} =1$
with an average
\begin{equation}
\label{norm.cond}
\langle ( \dots) \rangle_0 = \frac 1{N_0} \int \frac {d^3 k}{(2
  \pi)^3} \delta(\epsilon_{\bf k}) (\dots)
\end{equation}
defined on the Fermi surface.

In the BCS mean-field (MF) approximation the Hamiltonian is
diagonalized by the Bogoliubov transformation with parameters
\begin{eqnarray}
|u_{\bf k}|^2&=&(1+\epsilon_{\bf k}/E_{\bf k})/2 \\
|v_{\bf k}|^2&=&(1-\epsilon_{\bf k}/E_{\bf k})/2 \\
E_{\bf k} &=& \sqrt{\epsilon_{\bf k}^2 + |\Delta_0|^2 \psi^2({\bf k})}
\end{eqnarray}
The self-consistency equation for the MF approximation reads
\begin{equation}
\label{selfcon}
1 = V  \int' \frac{d^3k}{(2 \pi)^3} |\psi({\bf k})|^2
\frac{1-2f(E_{\bf k})}{2 E_{\bf k}}
\end{equation}
with the Fermi-Dirac distribution function $f(E)=1/[1+\exp(E/T)]$.
The primed integral runs over ${\bf k}$ with $|\epsilon_{\bf
  k}|<\epsilon_{\rm c}$ only.

For $\Delta_0 \to 0$ Eq. (\ref{selfcon}) implicitly determines
\begin{equation}
T_{\rm c}=1.14 \ \epsilon_{\rm c} \exp(-1/V N_0).
\end{equation}
Due to the choice of the normalization of $\psi$,
this relation does not depend explicitly on pairing symmetry
 or mass anisotropy.

In the critical region the amplitude of the gap follows from Eq.
(\ref{selfcon}) after expansion in second order of $\Delta_0$ and
linearizing in $1-T/T_{\rm c}$:
\begin{equation}
\label{gapcoeff}
  |\Delta_0(T)|^2 = \langle \psi^4 \rangle_0^{-1} \frac {8 \pi^2}{7 \zeta (3)}
(1-T/T_{\rm c}) \ T_{\rm c}^2 .
\end{equation}
In comparison to the isotropic $s$-wave case this result is modified
by a factor $\langle \psi^4 \rangle_0 = 15/7 $.

We now calculate for fixed electrochemical potential the change in
electron density due to formation of a gap. In the superconducting
state a single electron state ${\bf k}$ has occupation number
\begin{equation}
n^{\rm s}_{\bf k}=|v_{\bf k}|^2+(|u_{\bf k}|^2-|v_{\bf k}|^2)f(E_{\bf k})
\end{equation}
compared to the normal state with $n^{\rm n}_{\bf k}=f(\epsilon_{\bf
  k})$. The change of the total electron density
\begin{equation}
  \delta n  =  2\int' \frac{d^3k}{(2 \pi)^3}
\left(n^{\rm s}_{\bf k} -n^{\rm n}_{\bf k} \right)
\end{equation}
can be evaluated using the Sommerfeld approximation $N_\epsilon
\approx N_0+\epsilon N'_0$. One finds to lowest order in $\Delta_0$,
i.e. close to $T_{\rm c}$,
\begin{equation}
\label{del.n}
  \delta n \approx N'_0 |\Delta_0|^2 \ln \frac {\epsilon_{\rm c}}{T_{\rm c}},
\end{equation}
where again all dependence on pairing symmetry and mass anisotropy is
contained in $T_{\rm c}$. Since the condensate density $|\Delta_0|^2$
has a discontinuous slope near $T_{\rm c}$, Eq. (\ref{del.n}) implies
that for fixed electrochemical potential the electron density has a
discontinuous slope or, vice versa, the chemical potential for fixed
electron density. This effect and its consequences for the work
function has been pointed out by van der Marel\cite{workfun}.

\section{Method of classical trajectories}
\label{app.quasi}

As shown in Ref. \cite{SU91}, the kernel $K$ in Eq. (\ref{int.eqn}) is
determined by the following expression, which is valid for arbitrary
symmetry of the order parameter:
\begin{eqnarray}
\label{def.K}
 K({\bf r}_1,{\bf r}_2)&=&\pi VN_0T \sum _n\int_0^{\infty} dt~
  \exp(-2|\omega_n|t)
\nonumber \\
&& \times \left\langle \psi^*({\bf n}_1)
   \psi({\bf n}_2)\right\rangle _{\epsilon=0, {\rm classical}},
\end{eqnarray}
where $t>0$ is the time of motion along the trajectory and
$\omega_n=(2n+1)\pi T$ is a Matsubara frequency.  The angular brackets
denote the averaging over all possible classical trajectories ${\bf r}(t)$
of a particle, moving with the Fermi velocity $v_F$, which connect the
points ${\bf r}_1$ and ${\bf r}_2$ and satisfy the condition ${\bf r}(0)={\bf r}_1$
(one has to take into account both the trajectories, for which
${\bf r}(t)={\bf r}_2$, and the time-reversed trajectories, which give just a
Hermitian-conjugated contribution to the kernel). The unit vectors
${\bf n}_1$ and ${\bf n}_2$ denote the directions of velocity in the
corresponding points. We shall keep below the general notations for
the basis functions (which are chosen to be real), 
so that our results are applicable for any
one-dimensional spin-singlet order parameter.

Let us first consider the contribution from straight trajectories.
The expression in the angular brackets in the r.h.s. of (\ref{def.K})
has the form:
\begin{equation}
\label{av.direct}
 \langle (...)\rangle_{\rm direct}=\int\frac{d\Omega}{4\pi}~
  \delta({\bf r}_2-{\bf r}(t))\psi^2({\bf n})+{\rm h.c.},
\end{equation}
where ${\bf r}(t)={\bf r}_1+v_F{\bf n} t$ is the only straight path
from ${\bf r}_1$ to ${\bf r}_2$ (see Fig. \ref{fig.paths}a, ${\bf
  n}_1={\bf n}_2={\bf n}$). Here, integration is over all directions
of ${\bf n}$. Hence
\begin{equation}
  K_0({\bf r}_1-{\bf r}_2)=K_0({\bf r})= 
    \frac {V N_0}{4 \pi \xi_0} \frac {\psi^2({\bf n})}{r^2 \sinh(r/\xi_0)}
\end{equation}
with $\xi_0=v_{\rm F}/2\pi T_{\rm c}$, and ${\bf n}={\bf r}/|{\bf
  r}|$.  In the critical region the correlation lengths $\xi_i$ along
different directions are given by
\begin{equation}
\label{xis}
\xi^2_i(T) =\gamma_i\frac {7 \zeta (3)}{12} \left(1 -
  \frac {T}{T_{\rm c}}\right)^{-1} \xi_0^2.
\end{equation}
Due to the $d$-wave pairing, there is an anisotropy (even for the
spherical Fermi surface) described by the factors
$\gamma_X=\gamma_Y=9/7$ and $\gamma_Z= 3/7$.

The equation (\ref{gap.eqn}) follows from (\ref{int.eqn}) as a result of 
integration over the differences $y_2-y_1$ and $z_2-z_1$:
\begin{equation}
  \Delta(x_1)=\int_0^{\infty} dx_2~K(x_1,x_2)\Delta(x_2),
\end{equation}
the kernel being given by the sum of two terms:
\begin{equation}
 K(x_1,x_2)=K_0(x_1-x_2)+K_{\rm r}(x_1,x_2),
\end{equation}
where the contributions $K_0$ and $K_{\rm r}$ come respectively from
the straight trajectories and those ones, which are reflected against
the surface.

Substituting (\ref{av.direct}) in Eq. (\ref{def.K}), we obtain:
\begin{equation}
 K_0(x)=\frac{V N_0}{2 \xi_0} \sum _n\int_0^1 \frac{ds}{s}~
   F_0(s)\exp\left(-\frac{2|\omega_n|}{v_Fs}|x|\right),
\end{equation}
where
\begin{equation}
\label{F0}
 F_0(s)=\int_0^{2\pi}\frac{d\varphi}{4\pi}~\Bigl[
  \psi^2(s,\varphi)+\psi^2(-s,\varphi) \Bigr].
\end{equation}
Here $s=\cos\theta$, $\theta$ and $\varphi$ are the polar and
azimuthal angles respectively, so that ${\bf
  n}=(s,\sqrt{1-s^2}\cos\varphi,\sqrt{1-s^2}\sin\varphi)$.  The polar
axis is chosen along the normal vector to the surface (Fig.
\ref{fig.paths}). Note that $\int_0^1dsF_0(s)=1$ because of the
normalization condition of the basis functions.

The form of $K_{\rm r}(x_1,x_2)$ is determined by the reflection of
electrons at the boundary. For the case of specularly reflecting
boundary we have from (\ref{def.K}) the following expression:
$$
 \langle (...)\rangle_{\rm reflected}=\int\frac{d\Omega}{4\pi}~
  \delta({\bf r}_2-{\bf r}(t))\psi({\bf n}_1)\psi({\bf n}_2)+{\rm h.c.}
$$
The only possible reflected path is (see Fig.
\ref{fig.paths}b):
\begin{equation}
\label{tr.refl}
 {\bf r}(t)=\left\{
     \begin{array}{ll}
       {\bf r}_1+v_F{\bf n}_1t & \quad ,\quad t<t_0=-\frac{x_1}{v_Fs_1}~,\\
       v_F{\bf n}_2(t-t_0) & \quad ,\quad t>t_0~,
     \end{array} \right.
\end{equation}
where $t_0$ corresponds to the moment when the particle hits the
surface and ${\bf n}_{1,2}=(\pm s,\sqrt{1-s^2}\cos\varphi,
\sqrt{1-s^2}\sin\varphi)$ ($s_1<0$). Performing integrations we
obtain:
\begin{eqnarray}
 K_{\rm r}(x_1,x_2)=\frac{V N_0}{2 \xi_0} \sum _n\int_0^1
 \frac{ds}{s}~
   F_{\rm r}(s) \nonumber \\
  \times\exp\left(-\frac{2|\omega_n|}{v_Fs}(x_1+x_2)\right),
\end{eqnarray}
where
\begin{equation}
\label{Fspec}
 F_{\rm r}(s)=\int_0^{2\pi}\frac{d\varphi}{2\pi}~
  \psi(s,\varphi)\psi(-s,\varphi).
\end{equation}
The results (\ref{F0}) and (\ref{Fspec}) 
obviously coincide with those obtained in
Ref.\cite{Sam1} by another method.

In the case of diffusive surface one needs to take into account all
possible trajectories, reflected in all directions (see Fig.
\ref{fig.paths}c):
\begin{eqnarray*}
 \langle (...)\rangle_{\rm reflected}=\int\frac{d\Omega_1}{4\pi}
  \int d\Omega_2~P({\bf n}_1,{\bf n}_2)~\delta({\bf r}_2-{\bf r}(t))
\nonumber \\
 \times  \psi({\bf n}_1)\psi({\bf n}_2)+{\rm h.c.},
\end{eqnarray*}
where $P({\bf n}_1,{\bf n}_2)$ is the probability distribution for the
particle to be reflected in the direction ${\bf n}_2$, provided it
moved before along the direction ${\bf n}_1$. Assuming the reflection
is completely diffusive (i.e. the "memory" is completely lost), we
obtain that this probability density depends only on ${\bf n}_2$. As
usual, it is convenient to choose it in the following manner: $P({\bf
  n}_2)=\frac{1}{\pi}\cos \theta_2= \frac{1}{\pi}s_2$. The
corresponding trajectories are given by (\ref{tr.refl}), where ${\bf
  n}_1$ and ${\bf n}_2$ are now independent unit vectors.  After
integrations, the resulting expression for $K_{\rm r}$ is \cite{Sam2}:
\begin{eqnarray}
 K_{\rm r}(x_1,x_2)=\frac{V N_0}{2 \xi_0} \sum_n\int_0^1 ds_1
   \int_0^1 ds_2~F_{\rm r}(s_1,s_2)
\nonumber \\
  \times
    \exp\left(-\frac{2|\omega_n|}{v_F}\left(\frac{x_1}{s_1}+
     \frac{x_2}{s_2}\right)\right),
\end{eqnarray}
where
\begin{eqnarray}
\label{Fdiff}
 F_{\rm r}(s_1,s_2)=\int\limits_0^{2\pi}\frac{d\varphi_1}{2\pi}
    \int\limits_0^{2\pi}\frac{d\varphi_2}{2\pi}\Bigl[
 \psi(s_1,\varphi_1)\psi(-s_2,\varphi_2)
\nonumber \\
 +\psi(-s_1,\varphi_1)
  \psi(s_2,\varphi_2) \Bigr].
\end{eqnarray}

Using the specific angular dependence (\ref{def.psi}) of the order
parameter, the general expressions (\ref{F0}) and (\ref{Fdiff}) lead
to Eqs. (\ref{F0.dwave}) and (\ref{Fdiff.dwave}).

Note that if the contribution from the diffusively reflected
trajectories vanishes then the integral equation (\ref{gap.eqn}) 
can be solved exactly using the Wiener-Hopf method. This takes place for any
spin-triplet order parameter, in particular for $p$-wave order
parameter in the superfluid phases of $^3$He \cite{AdGR74}.

\end{multicols}


\begin{references}

\bibitem[*]{byline} Present address: Theory of Condensed Matter Group,
  Cavendish Laboratory, University of Cambridge, Madingley Road, 
  Cambridge CB3 0HE, UK. \\
  Permanent address: L. D. Landau Institute for Theoretical Physics, 
  Kosygina Str. 2, 117940 Moscow, Russia.\\

\bibitem{KK} D.I. Khomskii and F.V. Kusmartsev, Pis'ma v Zh. Eksp.
  Teor. Fiz. {\bf 54}, 150 (1991) [JETP Lett. {\bf 54}, 145 (1991)];
  Phys. Rev. B {\bf 46}, 14245 (1992).

\bibitem{KF95} D.I. Khomskii and A. Freimuth, Phys. Rev. Lett. {\bf
    75}, 1384 (1995).

\bibitem{B+96} G. Blatter, M. Feigel'man, V. Geshkenbein, A. Larkin,
  and A. van Otterlo, Phys. Rev. Lett. {\bf 77}, 566 (1996).

\bibitem{workfun} D. van der Marel, Physica C {\bf 165}, 35 (1990); G.
  Rietveld, N.Y. Chen, and D. van der Marel, Phys. Rev. Lett. {\bf
    69}, 2578 (1992).

\bibitem{Kri95} Cl.~Kriebel {\em et al}, Ann.~d.~Phys.~{\bf 4}, 136
  (1995).

\bibitem{d-wave} see, e.g., J. Annett, N. Goldenfeld, and A.J.
  Leggett, in {\it Physical Properties of High-Temperature
  Superconductors}, Vol. 5, ed. D.M. Ginsberg (World Scientific,
  Singapore, 1996).

\bibitem{d-vortex} G.E. Volovik, Pis'ma v Zh. Eksp. Teor. Fiz. {\bf 58},
  457 (1993) [JETP Letters {\bf 58}, 469 (1993)]; P.I. Soininen, C.
  Kallin, and A.J. Berlinsky, Phys. Rev. B {\bf 50}, 13883 (1994); R.
  Heeb, A. van Otterlo, M. Sigrist, and G. Blatter, Phys. Rev. B {\bf
    54}, 9385 (1996).

\bibitem{dG66} see, e.g., P.G. de Gennes, {\it Superconductivity of
  Metals and Alloys} (Benjamin, New York, 1966).

\bibitem{SU91} M. Sigrist and K. Ueda, Rev. Mod. Phys. {\bf 63}, 239
  (1991).

\bibitem{Luders} G. L\"uders and K.D. Usadel, {\it The Method of the
  Correlation Function in the Superconductivity Theory} (Springer,
  Berlin, 1971).

\bibitem{Sam1} K.V. Samokhin, Zh. Eksp. Teor. Fiz. {\bf 107}, 906 (1995)
  [JETP {\bf 80}, 515 (1995)].

\bibitem{Sam2} K.V. Samokhin, Ph.D. Thesis, L.D. Landau Institute
  (1995), unpublished.

\bibitem{Agter} D.F. Agterberg and M.B. Walker, Phys. Rev. B {\bf 53},
  15201 (1996).

\bibitem{note} Even if one considers diffusively reflecting surfaces
  at the top and the bottom of the film, the dipole moments are small
  compared to the inner surface since $\xi_\perp \ll \xi_\parallel$.
  In addition, the dipole moments at the top and at the bottom have
  opposite orientation and therefore give a negligible contribution to
  the lens effect.

\bibitem{dG64} P.G. de Gennes, Rev. Mod. Phys. {\bf 36}, 225 (1964).

\bibitem{CGM62} C. Caroli, P.G. de Gennes, and J. Matricon, J. Phys.
  Rad. {\bf 23}, 707 (1962); see also Ref. \CITE{dG66} p. 229.

\bibitem{DM87} G. Deutscher and K.A. M\"uller, Phys. Rev. Lett.  {\bf
    59}, 1745 (1987).

\bibitem{s+d} J.-H. Xu, Y. Ren and C.S. Ting, Phys. Rev. B {\bf 52},
  7663 (1995); J.J. Vicente Alvarez, G.C. Buscaglia and C.A. Balseiro,
  Phys. Rev. B {\bf 54}, 16168 (1996)

\bibitem{Gross90} R. Gross {\em et al}, Phys. Rev. Lett. {\bf 64}, 228 
     (1990)

\bibitem{BGZ95} Yu.S. Barash, A.V. Galaktionov and A.D. Zaikin, 
   Phys. Rev. B {\bf 52}, 665 (1995)

\bibitem{Sha85} E.A. Shapoval, Zh. Eksp. Teor. Fiz. {\bf 88}, 1073 (1985)
  [Sov. Phys. JETP {\bf 61}, 630 (1985)]

\bibitem{AdGR74} V. Ambegaokar, P.G. de Gennes and D. Rainer,
  Phys. Rev. A {\bf 9}, 2676 (1974).


\end{references}
\end{document}